\documentclass[sigconf]{aamas} 

\usepackage{enumitem}

\usepackage{balance} 
\usepackage{graphicx}
\usepackage{subcaption}
\usepackage{algorithm}
\usepackage{algpseudocode} 
\usepackage{float} 
\usepackage[table]{xcolor} 
\usepackage{orcidlink}
\definecolor{indigo}{RGB}{115,80,255}

\doi{EXII2056}

\makeatletter
\gdef\@copyrightpermission{
  \begin{minipage}{0.2\columnwidth}
   \href{https://creativecommons.org/licenses/by/4.0/}{\includegraphics[width=0.90\textwidth]{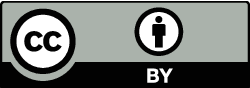}}
  \end{minipage}\hfill
  \begin{minipage}{0.8\columnwidth}
   \href{https://creativecommons.org/licenses/by/4.0/}{This work is licensed under a Creative Commons Attribution International 4.0 License.}
  \end{minipage}
  \vspace{5pt}
}
\makeatother

\setcopyright{ifaamas}
\acmConference[AAMAS '26]{Proc.\@ of the 25th International Conference
on Autonomous Agents and Multiagent Systems (AAMAS 2026)}{May 25 -- 29, 2026}
{Paphos, Cyprus}{C.~Amato, L.~Dennis, V.~Mascardi, J.~Thangarajah (eds.)}
\copyrightyear{2026}
\acmYear{2026}
\acmDOI{}
\acmPrice{}
\acmISBN{}

\acmSubmissionID{<<746>>}

\title[AAMAS-2026]{Adaptive Agents in Spatial Double-Auction Markets: Modeling the Emergence of Industrial Symbiosis}

\author{Matthieu Mastio\orcidlink{0009-0002-3486-3739}}
\affiliation{
  \institution{IRIT, UMR 5505 CNRS, Université Toulouse Capitole}
  \city{Toulouse}
  \country{France}
  \postcode{31000}
}
\email{matthieu.mastio@irit.fr}

\author{Paul Saves\orcidlink{0000-0001-5889-2302}}
\affiliation{
  \institution{IRIT, UMR 5505 CNRS, Université Toulouse Capitole}
  \city{Toulouse}
  \country{France}
  \postcode{31000}
}
\email{paul.saves@irit.fr}

\author{Benoit Gaudou\orcidlink{0000-0002-9005-3004}}
\affiliation{
  \institution{IRIT, UMR 5505 CNRS, Université Toulouse Capitole}
  \city{Toulouse}
  \country{France}
  \postcode{31000}
}
\email{benoit.gaudou@ut-capitole.fr}

\author{Nicolas Verstaevel\orcidlink{0000-0002-7879-6681}}
\affiliation{
  \institution{IRIT, UMR 5505 CNRS, Université Toulouse Capitole}
  \city{Toulouse}
  \country{France}
  \postcode{31000}
}

\email{nicolas.verstaevel@ut-capitole.fr}

\begin{abstract}
Industrial symbiosis fosters circularity by enabling firms to repurpose residual resources, yet its emergence is constrained by socio-spatial frictions that shape costs, matching opportunities, and market efficiency. Existing models often overlook the interaction between spatial structure, market design, and adaptive firm behavior, limiting our understanding of where and how symbiosis arises.
We develop an agent-based model where heterogeneous firms trade byproducts through a spatially embedded double-auction market, with prices and quantities emerging endogenously from local interactions. Leveraging reinforcement learning, firms adapt their bidding strategies to maximize profit while accounting for transport costs, disposal penalties, and resource scarcity. 
Simulation experiments reveal the economic and spatial conditions under which decentralized exchanges converge toward stable and efficient outcomes. Counterfactual regret analysis shows that sellers' strategies approach a near Nash equilibrium,  while sensitivity analysis highlights how spatial structures and market parameters jointly govern circularity.
Our model provides a basis for exploring policy interventions that seek to align firm incentives with sustainability goals, and more broadly demonstrates how decentralized coordination can emerge from adaptive agents in spatially constrained markets.
\end{abstract}

\keywords{Spatial Agent-Based Simulation, Industrial Symbiosis, Circular Economy, Distributed Market Analysis, Reinforcement Learning}

\newcommand{\BibTeX}{\rm B\kern-.05em{\sc i\kern-.025em b}\kern-.08em\TeX}

\begin{document}


\pagestyle{fancy}
\fancyhead{}


\maketitle 

\section{Introduction}

Industrial Symbiosis (IS) refers to the exchange of byproducts between firms, turning one company's waste into valuable inputs for another one. By reusing materials such as heat, rubble, or chemical residues, IS reduces environmental impact while generating economic benefits, embodying the circular economy principle of closing resource loops. For instance, the RETDA eco-industrial park in China reported a 33\% increase in input productivity, a 3650\% rise in water use efficiency, and a 30.91\% reduction in emissions \cite{chen2022}.

Despite its potential, IS rarely emerges spontaneously. Exchanges depend on local supply-demand compatibilities, economic incentives, and logistical constraints \cite{henriques2021industrial}. Poor coordination often results in wasted resources and missed opportunities, highlighting the gap between individual strategies and collective circularity goals.
%
Identifying the conditions under which decentralized firms can coordinate effectively is, therefore, a key research challenge.

Analytical solutions are often intractable due to the combinatorial complexity of interactions and the emergent nature of prices and traded quantities. Simulation-based approaches, by contrast, are particularly suited to study such nonlinear, decentralized, and heterogeneous systems \cite{he2019heterogeneous}. By systematically varying system configurations, simulations can indeed reveal market behaviors and highlight factors that support efficient and resilient local circularity. 
Agent-based models, in particular, are able to capture heterogeneity among actors and test how different market designs, decision rules, and learning mechanisms affect the emergence of exchanges under realistic economic constraints \cite{epstein1996growing}. Their ability to explicitly represent geosituated agents makes them especially suitable for studying spatially dependent interactions, such as proximity-based collaborations, transportation costs, and localized resource flows. Unlike theoretical game-theoretic models, which often rely on a small number of representative players to ensure tractability, agent-based models can accommodate many heterogeneous actors and interactions. 
i
However, their effectiveness is often constrained by the simplifications made in representing agent behaviors. In many existing symbiosis models \cite{Yazan2020,Ghali2017,Raimbault2020, Gan2015,FernandezMena2020,Chahla2019}, firms follow fixed behavioral rules or static pricing strategies, limiting their ability to capture the adaptive and uncertain nature of real-world exchanges. In practice, firms continuously adjust decisions in response to shifting demand, fluctuating costs, and evolving environmental regulations. This highlights the need to enhance decision-making mechanisms within agent-based models, enabling agents to learn, adapt, and explore strategies in a more dynamic way.

In this paper, we present a spatially explicit, decentralized multi-agent model in which buyers and sellers of byproducts interact through multilateral spatial double auctions using co-evolving strategies. Sellers adaptively update their pricing strategies via reinforcement learning in a partially observable environment, while buyers select offers that maximize utility subject to spatial and economic constraints. A key feature of the model is that both transaction prices and traded quantities emerge endogenously from these interactions, allowing us to investigate how local decisions give rise to systemic patterns.
While illustrated here through industrial symbiosis, this framework applies broadly to circular economy settings, such as heat recovery, wastewater reuse, or biomass valorization, where decentralized agents, ranging from firms to aggregated entities like neighborhoods, must coordinate to close resource loops under transport frictions. 
Our contributions are threefold. First, we develop a reinforcement learning-based market model for decentralized byproduct exchanges, based on a novel auction mechanism adapted to spatial and economic constraints of territorial circularity. Second, we validate the learned strategies via counterfactual regret analysis, showing convergence toward near-Nash equilibrium. Finally, we propose a simulation-based analysis of the conditions under which adaptive pricing and agent interactions lead to stable circular outcomes, providing insights for policymakers.

The paper first reviews related work (Section~\ref{sec:related}), then presents the decentralized market model (Section~\ref{sec:model}) and simulation setup (Section~\ref{sec:experiments}). Results on price formation and local circularity are analyzed in Section~\ref{sec:results}, followed by discussion (Section~\ref{sec:discussion}) and conclusions (Section~\ref{sec:conclusion}).

\section{Related Work}
\label{sec:related}

\subsection{Industrial Symbiosis and Circular Economy Modeling}
Industrial symbiosis in Eco-Industrial Parks (EIPs) has traditionally been studied using system dynamics to capture network level behavior and feedback loops~\cite{zhao2008simulation,qu2010system,sterman2000business,lorenz2006towards,cao2009applying}. Complementary approaches have treated EIPs as complex adaptive systems, highlighting dynamic business interactions and emergent symbiotic relationships~\cite{chertow2012organizing,yu2013process,felicio2016,lambert2002eco,romero2013framework}. Other studies focused on material and energy flows, byproduct exchanges, and social network analysis~\cite{Romero2014,felicio2016,yazan2016design,chertow2000,domenech2009social}. Recent reviews~\cite{Demartini2022,rizzati2024systematic} emphasize the potential of agent-based modeling to capture local, adaptive, and self-organizing behaviors. Emerging agent-based studies have integrated input-output models~\cite{Yazan2020}, explored spatially networked exchanges~\cite{Ghali2017,Raimbault2020, mastioCIEC}, and examined sector specific synergies~\cite{Gan2015,FernandezMena2020,Chahla2019}.
Yet, despite these advances, the explicit representation of economic incentives and market coordination mechanisms remains underexplored. 
Additionally, these works do not explicitly model markets where prices, trade volumes, and resource allocations emerge endogenously from decentralized interactions.

\subsection{Auctions and Market Mechanisms in Spatial Settings}
Auction and negotiation mechanisms in Multi-Agent Systems (MAS) have become canonical tools for decentralized coordination. 
From early negotiation platforms~\cite{Chavez1997} to double auctions—market mechanisms in which multiple buyers and sellers simultaneously submit bids and offers—used to study bidding strategies and price discovery~\cite{Tesauro2001,Satterthwaite2022}, research has demonstrated how decentralized negotiations can generate robust outcomes. Extensions have examined repeated competition~\cite{Ledvina2010,Bichler2024}, peer-to-peer trading~\cite{Qiu2021}, social influence~\cite{wainwright2024social}, and multilateral bargaining across multiple resources~\cite{Chen2015,An2011}.
Spatial aspects are frequently acknowledged, for instance through costly trade links~\cite{babaioff2005}, heterogeneous bilateral transport costs~\cite{chu2008}, or land markets where localized interactions induce differentiated prices~\cite{filatova2007}. Yet these contributions remain largely theoretical, typically involving few participants and limited attention to adaptive mechanisms that capture the evolving dynamics of real markets.

\subsection{Learning Agents for Adaptive Market Participation}
Traditional learning agents often rely on fixed rules~\cite{Cliff2024}, or supervised training~\cite{lauretto2013evaluation}, limiting their ability to adapt to dynamic market conditions and strategic interactions. Reinforcement Learning (RL) overcomes these limitations by allowing agents to learn optimal policies through trial-and-error, balancing exploration and exploitation in complex, evolving environments.
In the context of double auctions, RL agents are particularly well-suited due to the complex interactions between buyers and sellers and the continuous feedback provided by order book dynamics. Agents can be designed to adopt specific roles, such as deciding when to buy, sell, or set prices, enabling cooperative and strategic trading behaviors~\cite{bjerkoey2019, maeda2020}.
Multi-Agent Reinforcement Learning (MARL) extends this approach to settings with multiple interacting agents, allowing the simulation of bottom-up market dynamics~\cite{shavandi2022, karpe2020, lussange2020, ganesh2019}. MARL models have been shown to reproduce emergent behaviors in order books and centralized market microstructures by capturing interactions among heterogeneous agents~\cite{karpe2020}. This enables detailed analysis of price formation, liquidity, and strategic adaptation. 
However, existing approaches largely ignore spatial constraints, as these are typically absent in the stock market. Yet, such constraints can be critical in markets where transportation or delivery costs influence trading decisions~\cite{rodrigue2020geography}.
RL with spatial considerations has been primarily explored in logistics, where carriers and shippers act as bidders in dynamic auctions mediated by centralized brokers~\cite{Wang2018, Qiao2019, zha2017surge, atasoy2020}. Even when shippers themselves are modeled as learning agents~\cite{VanHeeswijk2019}, exchanges remain intermediary mediated. In contrast, many real-world IS transactions are inherently local and broker-free. To the best of our knowledge, RL has not yet been applied to decentralized, spatially constrained markets where transport costs and policy incentives influence agents' strategies. 

\subsection{Positioning of This Work}
Taken together, three main research gaps emerge:  
\begin{enumerate}[noitemsep, topsep=0pt] 
    \item \textbf{Industrial symbiosis models} rarely provide an in-depth mechanism for endogenous market-based price formation.  
    \item \textbf{Auction and negotiation models} seldom incorporate adaptive learning of agents.  
    \item \textbf{Reinforcement learning in markets} has not been applied to decentralized settings where spatial topology influences agents' behavior.  
\end{enumerate}

Our contribution is to bridge these strands by introducing a decentralized auction model for industrial symbiosis. While RL and double auctions have been studied in isolation, their integration into a spatial multilateral double-auction, where transport costs and location directly drive strategic behavior, constitutes a novel contribution. Specifically, our model ensures that (i) sellers adapt prices through reinforcement learning, (ii) spatial constraints are explicitly internalized, and (iii) both prices and traded quantities emerge from multilateral interactions. 
Our objective is to reproduce the systemic complexity of inter-firm interactions within industrial symbiosis as a subset of the broader circular economy, providing a foundational basis for modeling circular economic dynamics.
Beyond the specific application, the contribution lies in the rigorous methodology and implementation of this multi-agent RL framework, offering a reusable and generalizable approach for studying spatial market interactions.

\section{Model}
\label{sec:model}

\subsection{Environment and Agents} \label{env}
To concentrate on the essential dynamics of the model, we restrict our analysis to the exchange of a single byproduct. This simplification enables us to isolate and assess the effects of multilateral interactions and sellers' strategies, without additional complexity introduced by multiple interdependent byproducts.
We model a decentralized market for one byproduct, composed of buyers $\mathcal{B} = \{B_1, \dots, B_{N_B}\}$ and sellers $\mathcal{S} = \{S_1, \dots, S_{N_S}\}$. Each agent has a spatial location, and the distance between a buyer $B_i$ and a seller $S_j$ is denoted $d_{ij}$. Buyers have heterogeneous initial demands $q_i$ drawn from a scaled uniform distribution and evaluate incoming offers based on their price sensitivity $\beta_i$. Specifically, buyer $B_i$ accepts an offer $p_{ij}$ from seller $S_j$ only if $p_{ij} \le \beta_i \, p_m$, where $p_m$ is the reference market price. 
This mechanism ensures that buyers only engage in transactions that meet their cost expectations relative to the external market.
We adopt here a "Small Open Economy"\cite{GULATI2013288} assumption: we hypothesize that local industrial park internal exchanges are too small to impact the global commodity market price. Consequently, $p_m$ is treated as a fixed exogenous parameter.

Sellers are endowed with quantities $q_j$ and propose personalized prices to buyers, defined as
\begin{equation}
p_{ij} = \phi_j \  p_m + c_{ij},
\end{equation}
The term $c_{ij}$ is originally defined as $c_{ij} = d_{ij} \  c_t$,
where $c_t$ represents the transportation cost per kilometer (expressed as a percentage of the market price). This term can be further extended to capture the total logistics cost required for the byproduct provided by the seller to become usable by the buyer, including handling, treatment, or adaptation costs.  
Finally, $\phi_j$ is an adaptive pricing parameter that evolves during the learning phase. 

In addition to market and transportation costs, sellers face a penalty $c_d$ for unsold quantities (also expressed as a percentage of the market price), representing the cost of sending residual flows to landfill and reflecting both economic and environmental impacts. If a seller $j$ ends a timestep with $q_j^{\text{unsold}}$ units remaining, its reward is reduced by $q_j^{\text{unsold}} \  c_d$, imposing a direct financial penalty that encourages sellers to adjust pricing strategies to minimize waste while still aiming to maximize profit.

To characterize market conditions, we define the scarcity level of the byproduct as
\begin{equation}
s = \frac{\sum_{i \in \mathcal{B}} q_i}{\sum_{j \in \mathcal{S}} q_j}.
\end{equation}
This measure allows us to systematically investigate the effects of supply-demand imbalances on agents' behavior in the decentralized market.
Each seller $S_j$ aims to maximize its cumulative expected profit at each simulation step. 
Let $\mathbf{q}_j^{t} = (q_{ij}^{t})_{i \in \mathcal{B}}$ denote the vector of quantities bought by $B_i$ from $S_j$ at step $t$, and 
$\mathbf{p}_j^{t} = (p_{ij}^{t})_{i \in \mathcal{B}}$ the corresponding price vector. 
Defining $\mathbf{c}_j = (c_{ij})_{i \in \mathcal{B}}$ as the transport cost vector, the instantaneous profit can be written compactly as
\begin{eqnarray}
r_j^{t} = \langle \mathbf{q}_j^{t}, \, \mathbf{p}_j^{t} - \mathbf{c}_j \rangle \;-\; q_j^{\text{unsold},t} \  c_d ,
\end{eqnarray}
where $\langle \cdot, \cdot  \rangle$ denotes the dot product. 
The first term corresponds to the net revenue from executed contracts, while the second term penalizes unsold inventory through the landfill cost $c_d$.
This spatially defined market closely resembles a \emph{spatial Bertrand oligopoly} \cite{hotelling_stability_1929}, where multiple sellers compete by setting prices, and buyers prefer lower-cost options. 
Unlike the classical Bertrand model, where identical products drive prices toward marginal cost, the inclusion of transport costs differentiates sellers spatially, allowing local price variation and creating room for adaptive strategies to emerge. 

\subsection{Decentralized Multilateral Auction}
We model the interactions between buyers and sellers using a decentralized multilateral auction described in Algorithm \ref{algo}. Each seller submits a set of bids to each buyer, where each bid specifies a price and quantity for a given product $p$, computed on the basis of the logistic costs required to deliver the product to the buyer's location and the external market price.  
Buyers evaluate the incoming bids and select the one that maximizes their utility, subject to their price sensitivity $\beta_i$ and demand $q_i$. Selected bids are then sent back as proposals to sellers. Each seller chooses a subset of the received bids to accept in order to maximize expected profit, taking into account transport costs $c_{ij}$ and remaining inventory $q_j$. Contracts are executed, updating both buyers' needs and sellers' available quantities. This process iterates until no further mutually acceptable bids can be matched. Sellers adapt their pricing strategy $\phi_s$ over time using a strategy based on historical rewards that will be described in \ref{learning}.

\begin{algorithm}[ht] 
\caption{Decentralized Multilateral Auction}
\label{algo}
\begin{algorithmic}[1]
\Require Buyers $\mathcal{B}$, Sellers $\mathcal{S}$, product $p$
\While{contracts have been made at the previous timestep}
    \State \textbf{Seller bids:} each $S_j \in \mathcal{S}$ submits a bid $(q_j, p_{ij})$ to each $B_i \in \mathcal{B}$, where $q_j$ is the maximum quantity available for $S_j$, and $p_{ij} = \phi_j \  p_m + d_{ij} \  c_t$
    \State \textbf{Buyer choice:} each $B_i$ evaluates the received bids, selects the most attractive one within price tolerance $\beta_i p_m$, and sends an acceptance proposal to the corresponding seller
    \State \textbf{Seller allocation:} each $S_j$ sorts accepted bids by expected profit per sold unit, confirms contracts until inventory is exhausted, updates sold quantity and profit
    \State Remove satisfied buyers and empty sellers from $\mathcal{B}$ and $\mathcal{S}$
\EndWhile
\State \textbf{Return} executed contracts
\end{algorithmic}
\end{algorithm}

\subsection{Seller Reinforcement Learning and Action Selection} \label{learning}
It has been shown by \cite{Bichler2024} that Bertrand competition problems can be effectively learned using bandit algorithms. 
Accordingly, each seller’s sequential pricing decision is modeled as a contextual multi-armed bandit. 
The action space is discretized into $K$ intervals, $\forall \ 0<k \leq K$:
\begin{equation}
\Phi = \{\phi^{(1)}, \dots, \phi^{(K)}\}, \quad 
\phi^{(k)} = \phi_{\min} + \frac{k-1}{K-1}(1 - \phi_{\min}),
\end{equation}
where $\phi_{\min} = -\frac{c_d}{p_m}$. 
At every timestep $t$, seller $S_j$ selects a discrete pricing action $\phi_j^{t} \in \Phi$ based on the current market context and receives a reward $r_j^{t}$. 
Each interval is associated with a weight $w_k$ representing the expected reward.
When an action is chosen and the associated reward is obtained, the weights are updated via an exponential moving average (EMA):
\begin{eqnarray}
w_k^{(t+1)} = 
\begin{cases}
\alpha w_k^{t} + (1-\alpha) r_t, & \text{if } \phi_j = \phi^{(k)}, \\
w_k^{t}, & \text{otherwise}.
\end{cases}
\end{eqnarray}

Here, $\alpha \in [0,1]$ is the smoothing factor, which balances the influence of past and recent observations. By prioritizing recent data, EMA allows for rapid adaptation to changing conditions while remaining computationally efficient, requiring only constant-time updates per action. Although it does not offer formal convergence guarantees, EMA provides a practical compromise between adaptivity and scalability and allows for maintaining high learning efficiency \cite{morales2024exponential}.
Sellers select their next action using Boltzmann (softmax) sampling \cite{sutton1998reinforcement} with a declining temperature $\tau_t$:
\begin{equation}
P_k = \frac{\exp(w_k / \tau_t)}{\sum_{l=1}^{K} \exp(w_l / \tau_t)}, \quad
\tau_t = \max(\tau_{\min}, \tau_0 \cdot \text{decay}^t),
\end{equation}
where $\tau_0$ sets the initial exploration level, $\text{decay}$ controls the annealing rate, and $\tau_{\min}$ prevents premature convergence. This ensures exploration dominates early in learning, while the policy gradually becomes more conservative over time. The subsequent action $\phi_j^{(t+1)}$ is drawn at random from the set of actions according to probabilities $P_k$.
This sampling strategy balances exploration and exploitation in a multi-agent, stochastic environment. In our simulations, Boltzmann exploration consistently converged faster and with less oscillation than $\epsilon$-greedy or UCB, which often struggled with noise from multi-agent interactions~\cite{o2021variational}.
By combining a discretized action space, adaptive reward estimates, and Boltzmann sampling, the model captures the decentralized emergence of price formation and allocation dynamics in a spatially distributed market for byproducts. While Boltzmann does not always guarantee optimal regret bounds in multi-agent stochastic settings, its smooth adaptation makes it practical and effective for this scenario. 

\subsection{Symbiosis Index}
We finally introduce a \emph{symbiosis index} (SI) to quantify the efficiency of local byproduct exchanges. It is defined as:
\begin{equation}
\text{SI} = \frac{q_{\text{bought}}}{\min({q_{\text{toSell}}, q_\text{needed}})},
\end{equation}
where \(q_{\text{bought}}\) denotes the total quantity actually purchased, \(q_{\text{toSell}}\) the total quantity offered by the sellers, and \(q_{\text{needed}}\) the total quantity required by the buyers. The denominator normalizes the indicator so that it cannot exceed 1, even when the sold quantity is greater than the buyer’s need. This ensures that the symbiosis index reflects the proportion of demand satisfied rather than rewarding oversupply.
In contrast to existing technical indicators \cite{hardy2002industrial,tiejun2010two,zhao2008simulation,gao2013study}, which often incorporate numerous parameters (e.g., material compatibility, life cycle impacts, or multi-criteria environmental scores), our simplified formulation deliberately focuses only on exchanged quantities. This avoids unnecessary complexity while remaining sufficient for capturing efficiency in the present modeling framework.

\section{Experiments}
\label{sec:experiments}
\subsection{Simulation Framework}

The model is implemented in a simulation framework that supports decentralized markets with heterogeneous agents. Buyers and sellers are spatially distributed across a territory, which can be represented either with an abstract 2D environment or with a real geographic area with road network constraints. Distances, travel times, and transport costs are computed based on the underlying spatial representation, allowing realistic modeling of logistical constraints.  
The businesses interact through the auction mechanism described earlier, with offers, acceptances, and executed contracts dynamically updating inventories and demands. 

The framework tracks a variety of systemic metrics over time, including the local prices and their evolution, the proportion of demand satisfied locally (symbiosis index) and the trading patterns.
These indicators form the backbone of our analysis, as they connect local agent decisions to measurable systemic dynamics.  
This flexibility enables us to investigate a wide range of experimental settings, from dense and highly interconnected territories to sparse environments where exchanges are scarce. Visualization tools provide spatial and temporal representations of market dynamics, price formation, and circularity indicators.  
The code implementation is open source\footnote{\url{https://github.com/matthieu-mastio/MASTIO}}.

\subsection{Experimental Conditions}
\begin{figure}[htb]
    \centering
    \begin{subfigure}[b]{0.23\textwidth}
        \centering
        \includegraphics[width=\textwidth]{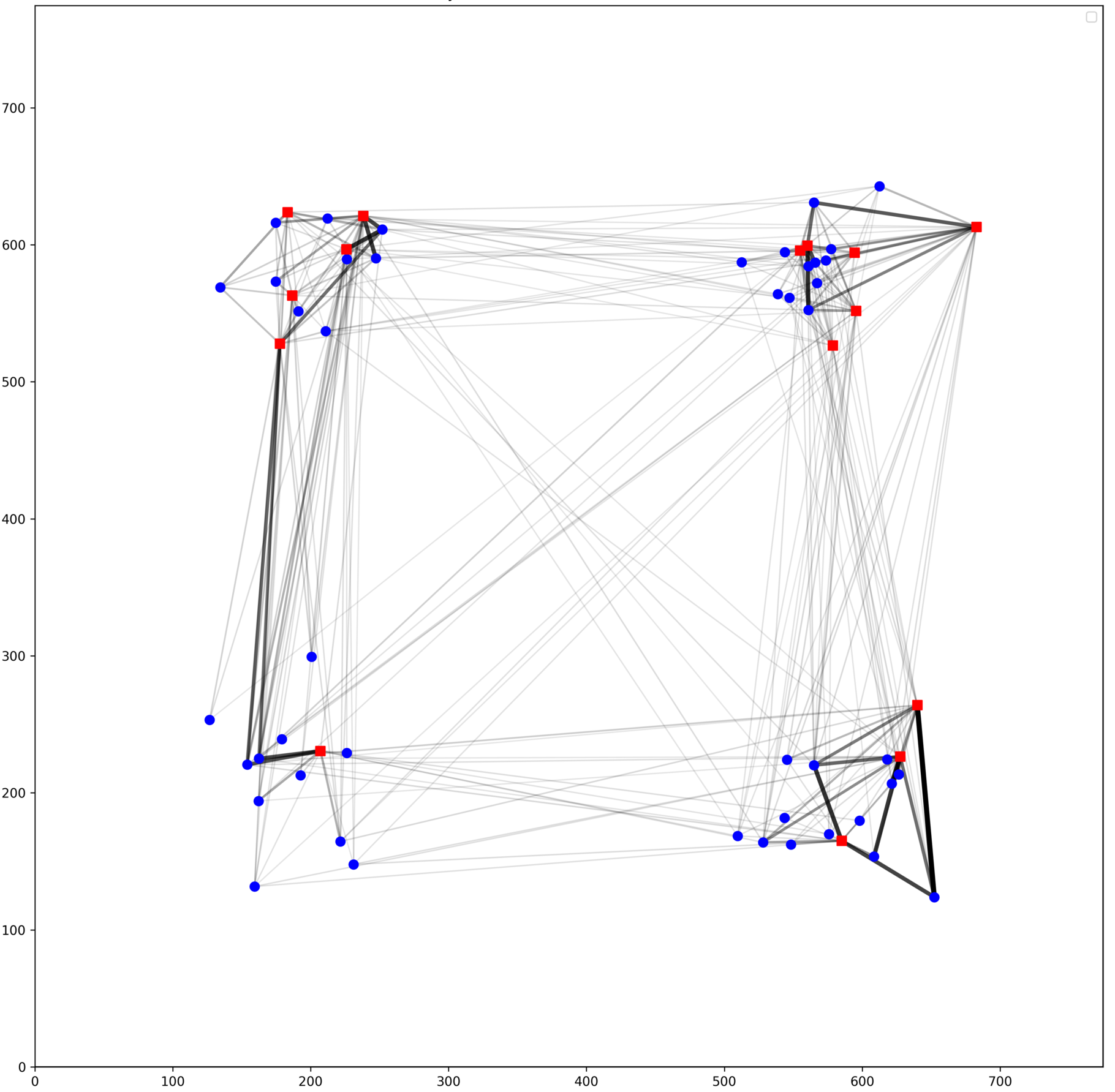}
        \caption{$cs = 0.05$}
        \label{fig:clust005}
    \end{subfigure}
    \hfill
    \begin{subfigure}[b]{0.23\textwidth}
        \centering
        \includegraphics[width=\textwidth]{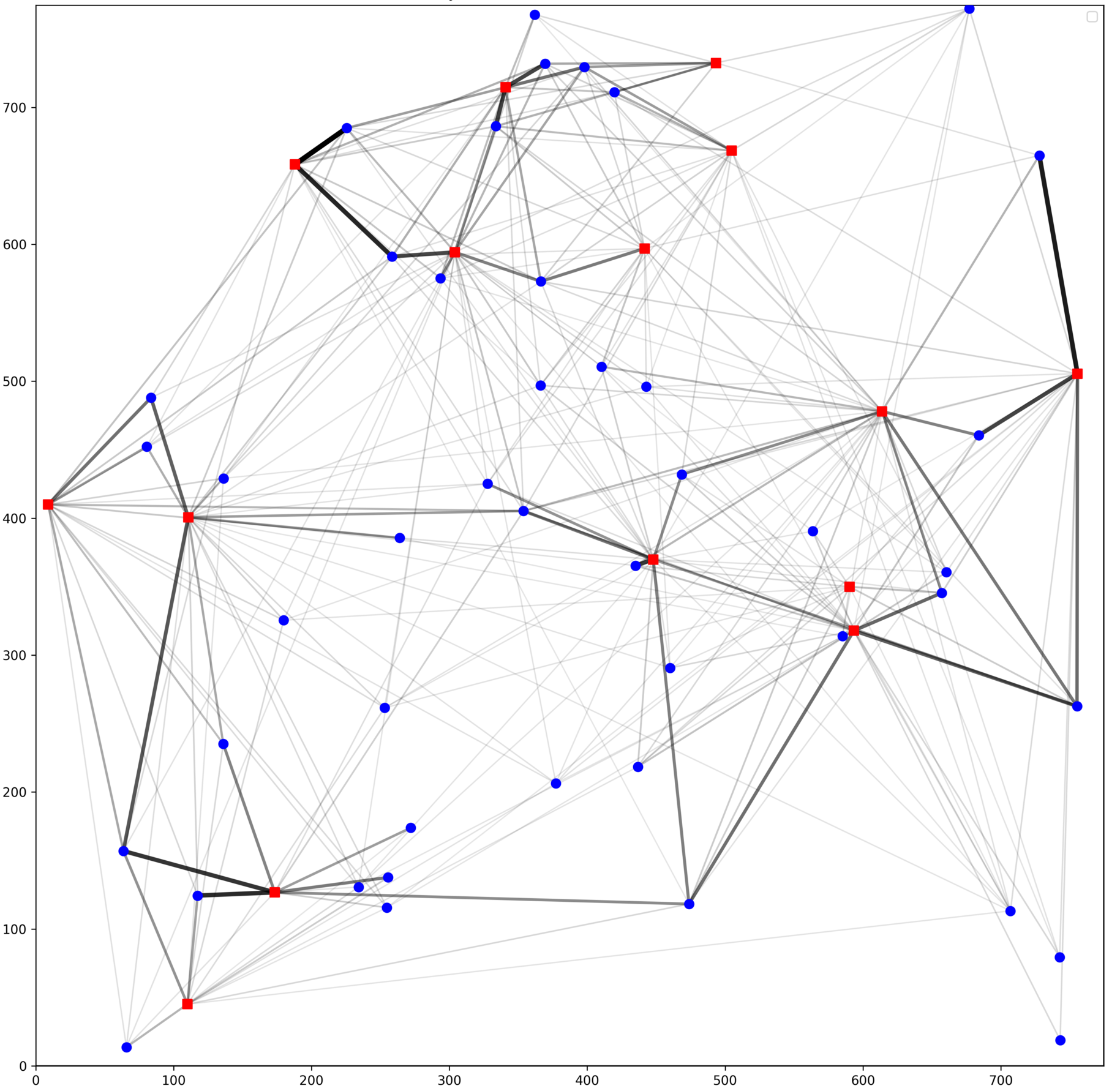}
        \caption{$cs= 0.5$} 
        \label{fig:clust05}
    \end{subfigure}
    \caption{Comparison of spatial agent distributions for different cs values.}
    \Description{Comparison of spatial agent distributions for different cluster spread values: 2 cases are displayed, the first one with cs = 0.05, showing 4 dense clusters; the second one with cs = 0.5, displaying an almost uniform distribution of agents in space.}
    \label{clust}
    \vspace{-0.3cm} 

\end{figure}

For similar reasons behind our decision to focus on a single product, we decided to focalize on virtual environments. This approach provides precise control over key parameters such as firm density, territorial extent, and spatial clustering, thereby enabling a systematic exploration of their influence on circularity indicators and overall market performance.
One of the key factors shaping the potential for cooperation is the spatial dispersion of firms.
We therefore define the cluster spread parameter ($cs$) as the standard deviation of companies' positions around cluster centers, normalized by the environment width:
small values lead to compact, well-separated clusters, while setting it to $1$ produces a configuration comparable to a fully uniform distribution across the environment, effectively removing any clustering effect. Figure~\ref{clust} shows the difference between a low (\ref{fig:clust005}) and high (\ref{fig:clust05}) spread value.
In our experiments, businesses were allocated into four spatial clusters. By default, the cluster spread parameter was set to $1$, indicating a broad dispersion and minimal clustering.

%
The results we obtained came from simulations conducted with a population of 40 firms, representing the typical scale of a standard EIP. 
The market price was fixed at $100$, and the transportation cost was set to $0.1$ per kilometer. 
This transport cost is a normalized parameter relative to the market price, calibrated to ensure that spatial distance acts as a meaningful economic friction without completely blocking trade, a choice further supported by the sensitivity analysis in Section \ref{sa}.
Sellers adjusted their strategies through Boltzmann exploration, with a temperature decay parameter set to $0.996$ and the action space discretized into $30$ bins.
The relatively small scale of the simulated market was chosen to balance computational tractability with the ability to capture complex multi-agent interactions. This size of population allows us to conduct thousands of simulation runs for systematic sensitivity analyses and counterfactual experiments, which would be computationally prohibitive at larger scales. Furthermore, each simulation runs for a large number of timesteps (1000 steps) to ensure that seller strategies have sufficient time to adapt and converge, and that emergent patterns of price formation and allocation stabilize. By prioritizing repeated experimentation and long-horizon dynamics over absolute market size, we focus on understanding the fundamental mechanisms driving spatial effects, adaptive bidding behavior, and the emergence of circularity.
 Our implementation is optimized for performance, allowing on a standard laptop processor (Intel(R) Core(TM) Ultra 9 185H processor with 21 cores) to achieve approximately 100 simulations of 1000 timesteps per minute.

\section{Results}
\label{sec:results}
In this section, we examine the core dynamics of the decentralized, spatially explicit market, focusing on conditions for price convergence and the efficiency of adaptive strategies using counterfactual regret analysis. We then explore how spatial organization, resource scarcity, disposal costs, and density interact to shape market outcomes and local circularity. Finally, a global sensitivity analysis quantifies the influence of individual parameters and their interactions on emergent symbiosis patterns.

\subsection{Convergence Toward Equilibrium}
\begin{figure}[htb]
    \centering
    \begin{subfigure}[b]{0.495\linewidth}
        \centering
        \includegraphics[height=3.8cm,width=1.03\linewidth]{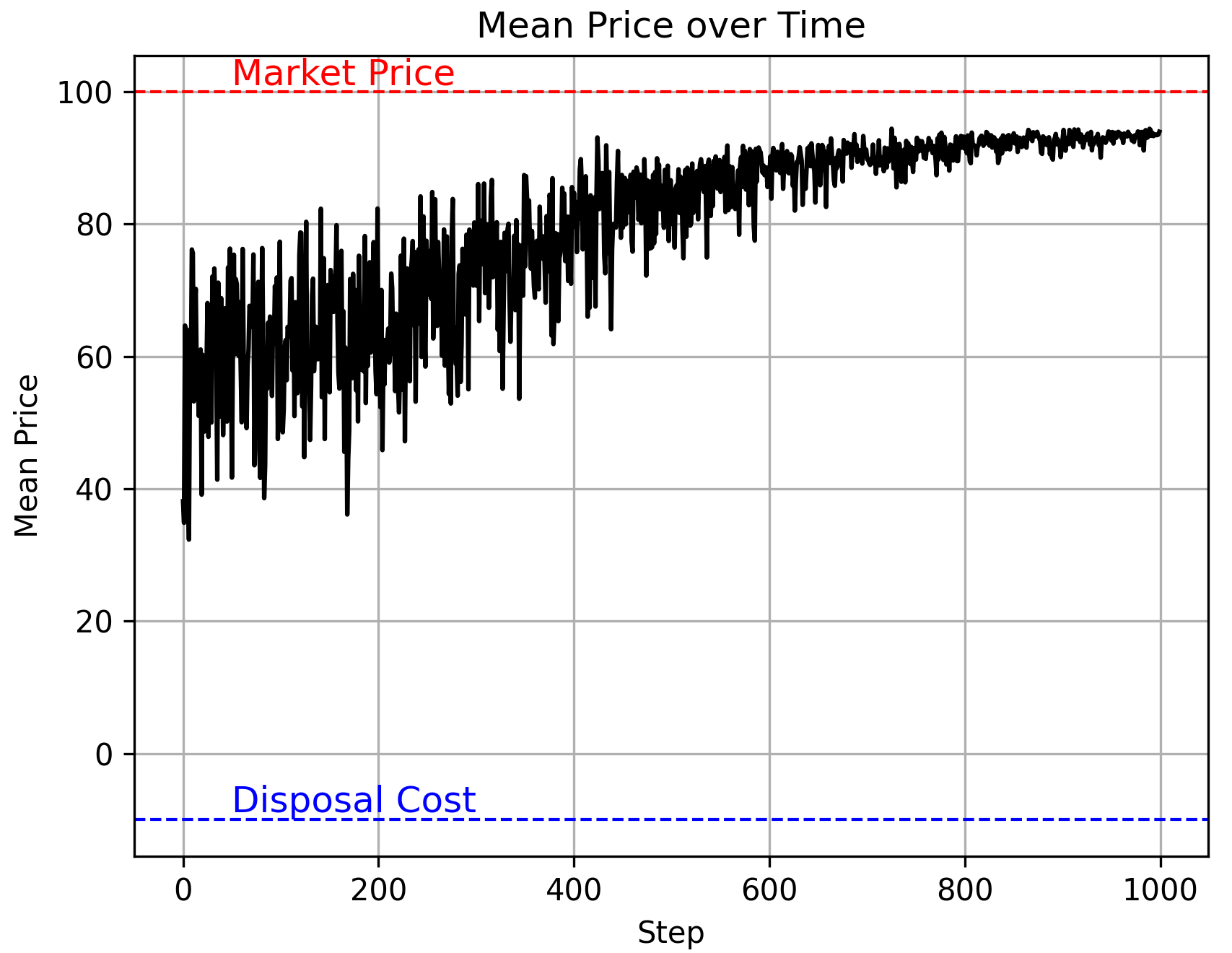}
        \caption{$s=2$, $c_d=10$}
        \label{fig:mean_price_high_scarcity}
        \Description{Mean transaction price trajectory for high scarcity ($s=2$) and low disposal cost ($c_d=10$).}
    \end{subfigure}
    \hfill
    \begin{subfigure}[b]{0.495\linewidth}
        \centering
        \vspace{-0.15cm}
        \includegraphics[height=3.8cm,width=1.03\linewidth]{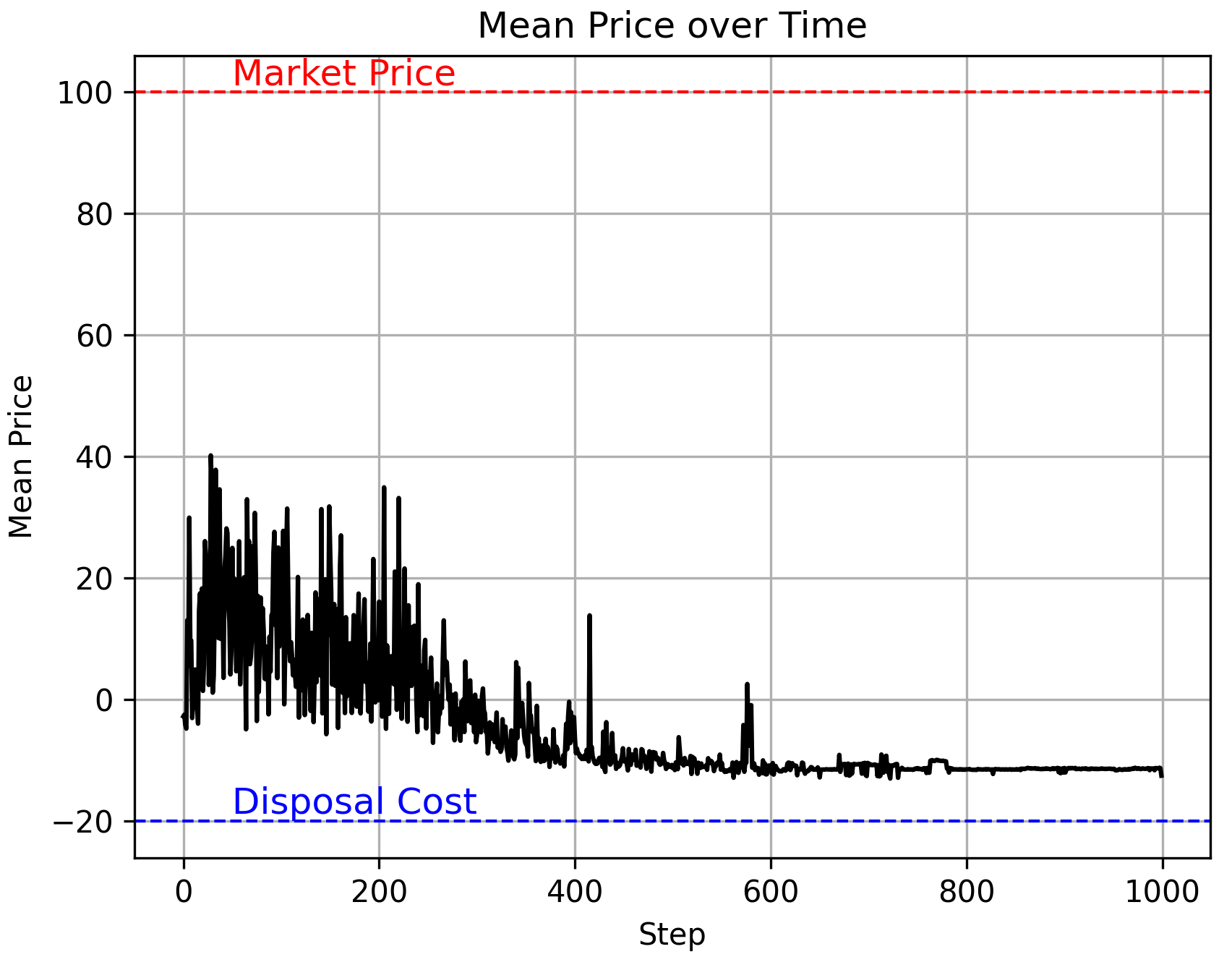}
        \caption{$s=0.5$, $c_d=20$}
        \label{fig:mean_price_low_scarcity}
        \Description{Mean transaction price trajectory for low scarcity ($s=0.5$) and high disposal cost ($c_d=20$).}
    \end{subfigure}
    \caption{Evolution of mean transaction prices under different scarcity and disposal cost conditions, where $s$ denotes the scarcity ratio and $c_d$ the disposal cost.}
    \label{fig:mean_price_comparison}
        \vspace{-0.3cm} 

\end{figure}

We first explore empirically the evolution of transaction prices as sellers adapt their strategies 
under different scarcity and disposal cost conditions, with a density fixed to 0.001. 
Figures~\ref{fig:mean_price_high_scarcity} and~\ref{fig:mean_price_low_scarcity} 
show the mean price trajectories over time compared to the external market price and the disposal cost. 
As anticipated given the decaying temperature, the curves gradually stabilize, approaching an equilibrium after sufficient simulation steps.
In the high scarcity and low disposal cost setting (Figure~\ref{fig:mean_price_high_scarcity}), 
prices converge toward an equilibrium close to the external market price.  
By contrast, in the low scarcity and medium disposal cost scenario (Figure~\ref{fig:mean_price_low_scarcity}), 
prices decline steadily and stabilize near the disposal cost, reflecting buyers’ stronger bargaining power and the reduced competitive pressure on resource demand.  

To better understand the efficiency and stability of the sellers' adaptive strategies, we next examine the dynamics at the level of individual firms using counterfactual regret analysis \cite{zinkevich2007regret}, which quantifies how much a seller could have improved its payoff by choosing alternative actions.

Let \(\phi_j^{(t)} \in \Phi_j\) denote the action used by seller $j$ at timestep $t$, and  \(r_j^{(t)}(\phi_j^{(t)})\) denote the reward received at that step.  
The \emph{optimal reward} at time $t$ is
\begin{equation}
r_j^{\star (t)} = \max_{\phi \in \Phi_j} r_j^{(t)}(\phi),
\end{equation}
where \(r_j^{(t)}(\phi)\) is the reward that the seller $j$ would have obtained at step $t$ if it had chosen action \(\phi\) while all other sellers' actions remained unchanged.
The \emph{regret} of seller $j$ at time $t$ is then
\begin{equation}
\text{Regret}_j^{(t)} = r_j^{\star (t)} - r_j^{(t)} \left(\phi_j^{(t)} \right).
\end{equation}
A positive regret indicates that the seller could have achieved a higher payoff at that step by selecting a different action, whereas a regret of zero indicates that the chosen action was optimal given the choices of the other sellers at that moment.
\begin{figure}[htb]
        \centering
        \includegraphics[height=5cm,width=0.8\linewidth]{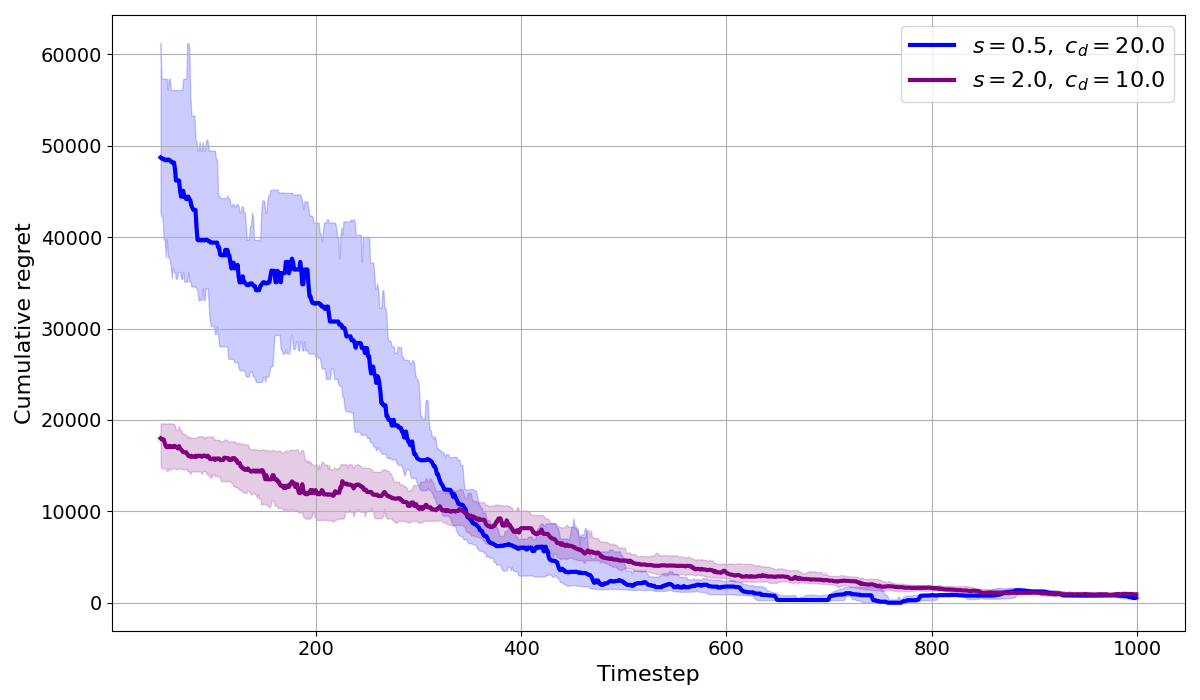}

        \Description{Rolling median of per-step regret for high scarcity ($s=2$) and low disposal cost ($c_d=10$).}
    \caption{Rolling median ($window=50$) of per-step regret under different scarcity and disposal cost conditions.}
    \label{fig:regret_comparison}

  \vspace{-0.4cm}
\end{figure}

Empirically, we observe that the per-step total regret, the aggregate deviation from a joint optimum, declines over time and converges toward zero in both scenarios (Figure~\ref{fig:regret_comparison}).
Since per-step regret measures the immediate payoff loss relative to the best fixed action in hindsight at that step, a vanishing regret implies that sellers' strategies are nearly optimal given the current actions of others. 
Consequently, the joint strategy profile of all sellers appears to converge toward an approximate \emph{Nash equilibrium}, providing justification for the stability of the emergent market dynamics under our MARL learning scheme.

\subsection{Price and Symbiosis Dynamics}
\begin{figure}[htb]
  \centering
  \includegraphics[height=3.8cm, width=\linewidth]{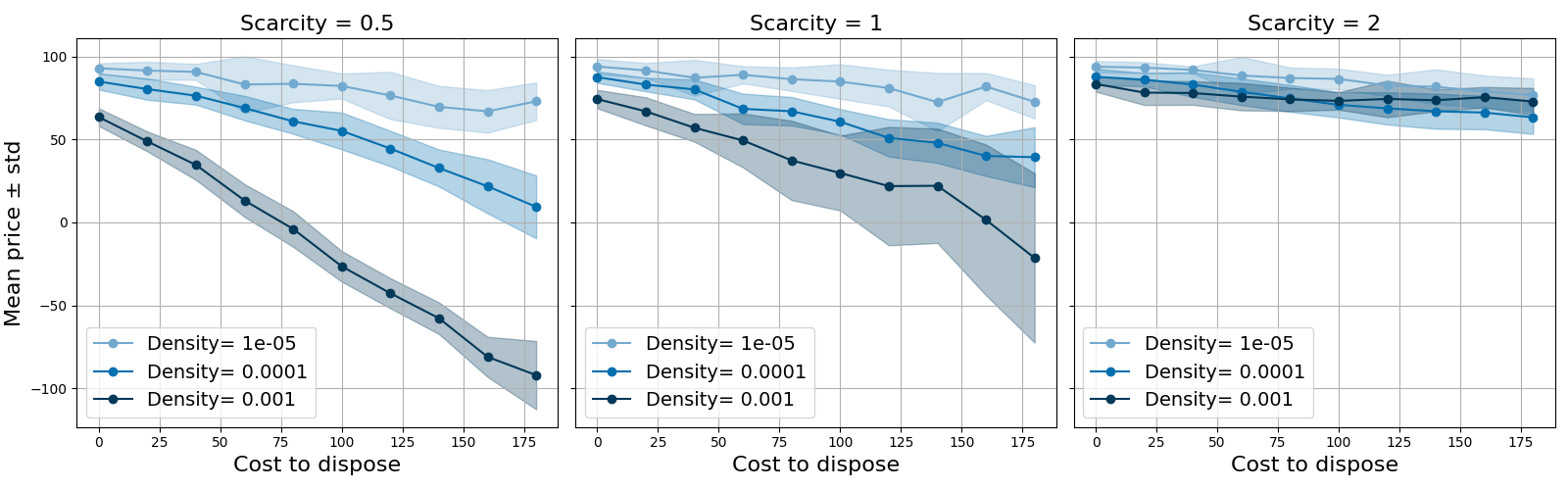}
  \caption{Evolution of the exchange price}
  \label{fig:price}
  \Description{Evolution of the exchange price}
  \vspace{-0.2cm}
\end{figure}

\begin{figure}[htb]
  \centering
  \includegraphics[height=3.8cm, width=\linewidth]{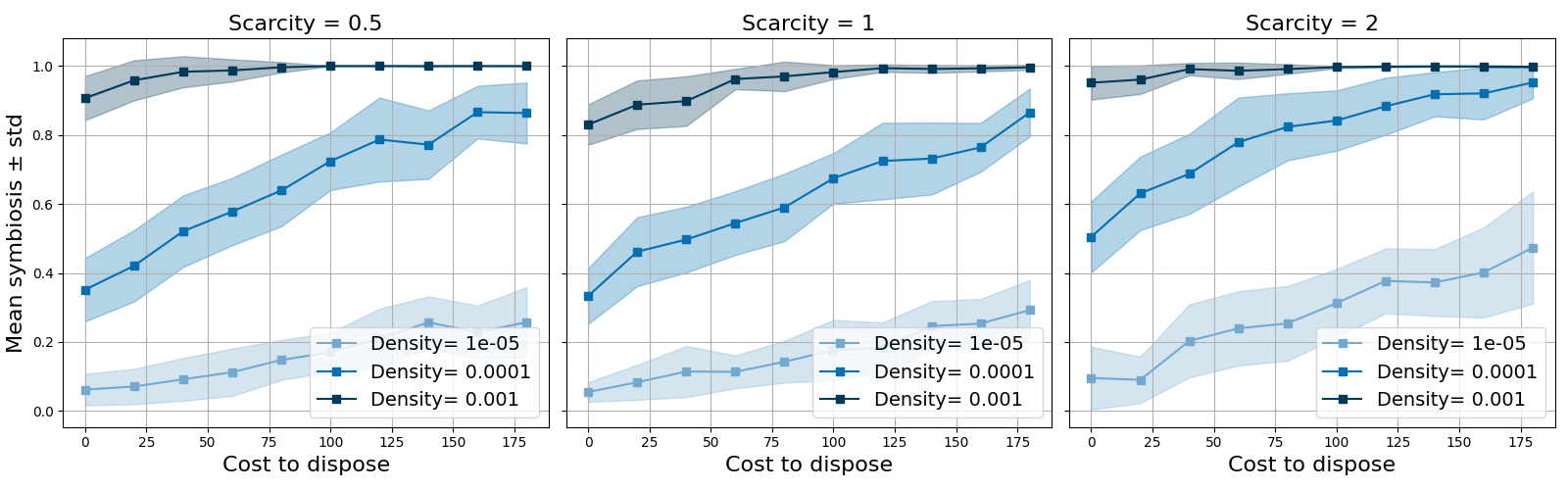}
  \caption{Evolution of symbiosis index}
  \label{fig:sym}
  \Description{Evolution of symbiosis index}  
  \vspace{-0.4cm}
\end{figure}

In this section, we examine how firm density, resource scarcity, and disposal costs shape the dynamics of transaction prices and circularity.
We focus on these output metrics because they directly quantify the level of local circularity achieved in the market and the average exchange price, making them relevant proxies for assessing the effectiveness of different policy levers.
By systematically varying firm density, resource scarcity, and disposal costs, we analyze how structural and economic factors influence sellers’ adaptive strategies, shaping both the emergence of local symbiosis and overall market outcomes.
Figures~\ref{fig:price} and~\ref{fig:sym} illustrate the evolution of equilibrium prices and local symbiosis levels as a function of disposal cost, scarcity, and density. Each point is the mean of 10 simulation runs, with shaded areas indicating one standard deviation to visualize variability across runs.

On the price side (Figure~\ref{fig:price}), we observe that, as expected, higher disposal costs induce sellers to lower their prices in order to avoid the landfill penalty. This effect is strongest in low scarcity environments ($s=0.5$), where intense competition can even drive equilibrium prices below zero in dense territories. At intermediate scarcity ($s=1$), the downward trend remains but is attenuated, as sellers retain partial market power. Under high scarcity ($s=2$), prices remain largely insensitive to $c_d$: the excess demand ensures that buyers continue to pay close to the outside option $p_m$, leaving sellers with little incentive to adjust their strategies.
Interestingly, at the demand equilibrium ($s=1$), we observe higher variance, reflecting the instability and potential oscillations in companies’ strategies as they adapt to competitive pressures.

Symbiosis levels (Figure~\ref{fig:sym}) exhibit a complementary pattern: the share of demand met through local exchanges increases with $c_d$, as the disposal penalty strengthens the incentive to form local contracts. Here, density plays a critical role: in dense environments, local exchanges dominate rapidly, and symbiosis saturates close to one; in sparse environments, by contrast, symbiosis remains low even under high $c_d$, reflecting structural limitations in firms' connectivity. 
We observe that this time scarcity has only a marginal effect on the evolution of symbiosis: higher scarcity slightly enhances local exchanges, but the dominant factors remain disposal cost and firm density.


\subsection{Sensitivity Analysis}
\label{sa}
The previous analysis showed that market outcomes are shaped by interacting mechanisms, but it did not allow us to fully disentangle their relative influence. To address this limitation and account for the influence of additional variables, we conducted a variance-based global sensitivity analysis~\cite{saltelli2010variance}.

We evaluate the influence of the simulation parameters on the symbiosis index and exchanged prices measured at the end of each simulation run after 1000 timesteps. 
By analyzing which parameters exert the greatest influence on the price and symbiosis index, we can identify the factors that policymakers should prioritize when designing interventions to foster circularity while keeping prices at a low level. We simulate varying values for the following five variables:

\begin{itemize}
    \item \textbf{Disposal cost} ($c_d \in [0,200]$), representing the penalty for unsold byproducts;
    \item \textbf{Scarcity level} ($s \in [0.25,2]$), controlling the relative abundance of resources in the market;
    \item \textbf{Firm density} ($\rho \in [10^{-5},10^{-1}]$), defining the number of firms per unit area;
    \item \textbf{Cluster spread } ($cs \in [0,0.5]$), determining how tightly companies are spatially grouped within clusters;
    \item \textbf{Transportation cost per kilometer} ($c_{t} \in [0,10]$), capturing the economic friction associated with distance.
\end{itemize}
 
To efficiently explore the parameter space, we employed surrogate models, which act as fast emulators of the full simulation while capturing the key input-output relationships~\cite{saves2024smt}. Specifically, we implemented a sparse Polynomial Chaos Expansion (PCE), enabling the analytical derivation of Sobol’ indices from its coefficients~\cite{sudret2008global}. The surrogate models demonstrated strong predictive performance, achieving an $R^2$ of 95\% for the symbiosis and 92\% for the price on a 20\% test sample.
Parameter sampling was performed using Latin hypercube sampling to train surrogate models, with $s$ sampled on a $\log_2$ scale and ${density}$ on a $\log_{10}$ scale.
For each parameter set, two independent simulations were executed to account for stochastic variability in agent decisions, resulting in a total of 20,000 independent simulations. 
Sobol’ sensitivity analysis decomposes the variance of the model output into orthogonal contributions attributable to individual input parameters and their interactions. We computed the full set of Sobol’ indices: first-order indices quantify the direct effect of each parameter on output variance, while higher-order indices capture interaction effects among parameters~\cite{chastaing}. Taken together, Sobol’ indices provide a rigorous and comprehensive assessment of the relative importance of independent inputs and their interactions, with their sum accounting for $100\%$ of the variance.

In Table~\ref{tab:sobol_symb_price_top}, out of 31, we show only the 12 indices greater than $1\%$ of the total variance for either symbiosis or price indices.
The Sobol' analysis confirms the strong influence of firm density: higher density reduces the average distance between agents and intensifies competition, explaining its impact on both model outcomes. As expected, price is also strongly affected by scarcity, and both outputs are influenced by transport costs. Interestingly, higher-order interactions contribute substantially to the variance, emphasizing the importance of interplay effects between parameters.

\begin{table}[htb]
\centering
\caption{Comparison of Sobol' sensitivity indices for the \textit{symbiosis} and \textit{price} outputs, including unaccounted variance from Sobol’ indices < 1\%.}
\label{tab:sobol_symb_price_top}
\resizebox{\columnwidth}{!}{%
\begingroup
\tiny
\begin{tabular}{l|p{1.5cm}|p{1.5cm}}
\toprule
\multicolumn{1}{c|}{\textbf{Variables}} & \multicolumn{1}{c}{\textbf{Price Sobol'}} & \multicolumn{1}{c}{\textbf{Symbiosis Sobol'}} \\
\midrule
$\rho$          & \cellcolor{blue!25!indigo!30}$S_1 = 20.5\%$  & \cellcolor{blue!30!indigo!40}$S_1 = 63.2\%$ \\
$s$             & \cellcolor{blue!25!indigo!35}$S_1 = 24.6\%$  & \cellcolor{blue!10!indigo!6}$S_1 = 1.3\%$   \\
$c_{t}$         & \cellcolor{blue!15!indigo!13}$S_1 = 8.5\%$   & \cellcolor{blue!20!indigo!21}$S_1 = 16.7\%$ \\
$s ,\rho$       & \cellcolor{blue!20!indigo!20}$S_2 = 14.0\%$  & \cellcolor{white}$S_2 \approx 0$            \\
$c_d$           & \cellcolor{blue!20!indigo!17}$S_1 = 11.0\%$  & \cellcolor{blue!10!indigo!8}$S_1 = 2.4\%$   \\
$\rho , c_{t}$  & \cellcolor{white}$S_2 \approx 0$             & \cellcolor{blue!15!indigo!15}$S_2 = 7.5\%$  \\
$s , c_{t}$     & \cellcolor{blue!15!indigo!15}$S_2 = 7.0\%$   & \cellcolor{white}$S_2 \approx 0$            \\
$cs$            & \cellcolor{blue!10!indigo!6}$S_1 = 1.5\%$    & \cellcolor{blue!15!indigo!15}$S_1 = 3.8\%$  \\
$c_d, s$        & \cellcolor{blue!15!indigo!15}$S_2 = 4.6\%$   & \cellcolor{white}$S_2 \approx 0$            \\
$c_d, \rho$     & \cellcolor{blue!10!indigo!7}$S_2 = 1.9\%$    & \cellcolor{white}$S_2 \approx 0$            \\
$\rho , cs$     & \cellcolor{white}$S_2 \approx 0$             & \cellcolor{blue!15!indigo!15}$S_2 = 2.2\%$  \\
$c_d, s , \rho$ & \cellcolor{blue!10!indigo!6}$S_3 = 1.7\%$    & \cellcolor{white}$S_3 \approx 0$            \\
\midrule
\textbf{Unaccounted} & \cellcolor{blue!5!white}$\approx 4.7\%$ & \cellcolor{blue!10!white}$\approx 3.0\%$ \\
\bottomrule
\end{tabular}
\endgroup
}
\end{table}

\begin{figure*}[!htb]
    \centering
    \begin{subfigure}[b]{0.45\linewidth}
        \centering
        \includegraphics[height=6.25cm,width=\linewidth]{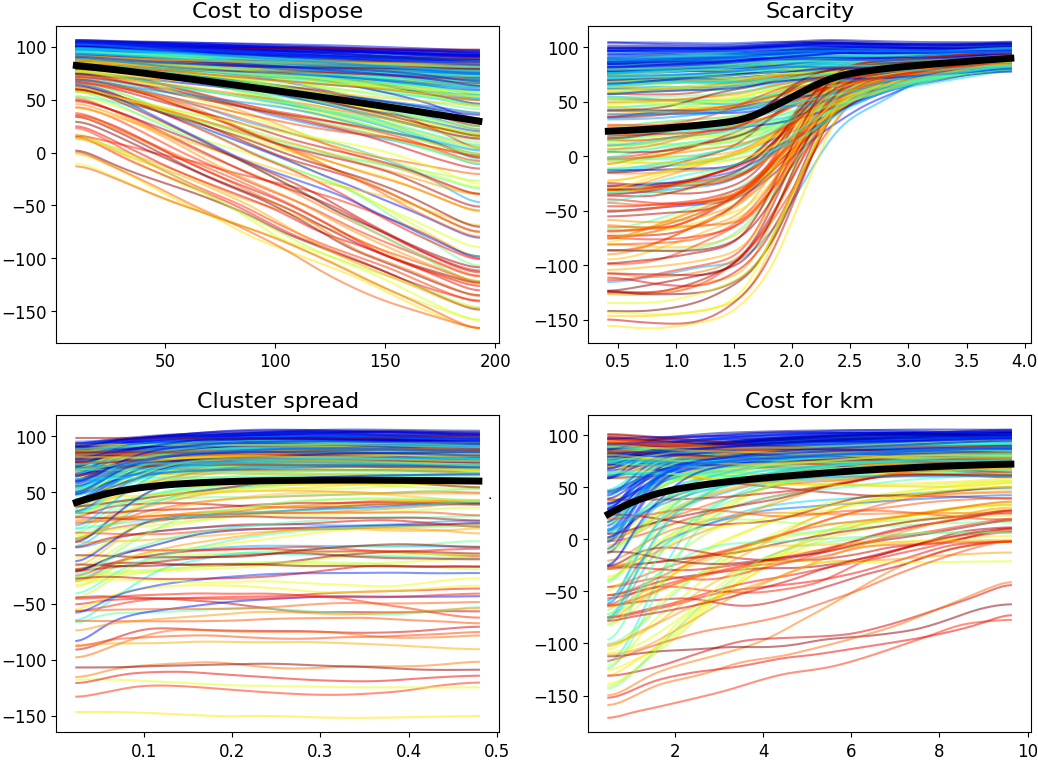}
        \caption{PDP and ICE illustrating the factors influencing the price index}
        \label{fig:pdp_price}
    \end{subfigure}
    \hfill
    \begin{subfigure}[b]{0.45\linewidth}
        \centering
        \includegraphics[height=6.25cm,width=1.05\linewidth]{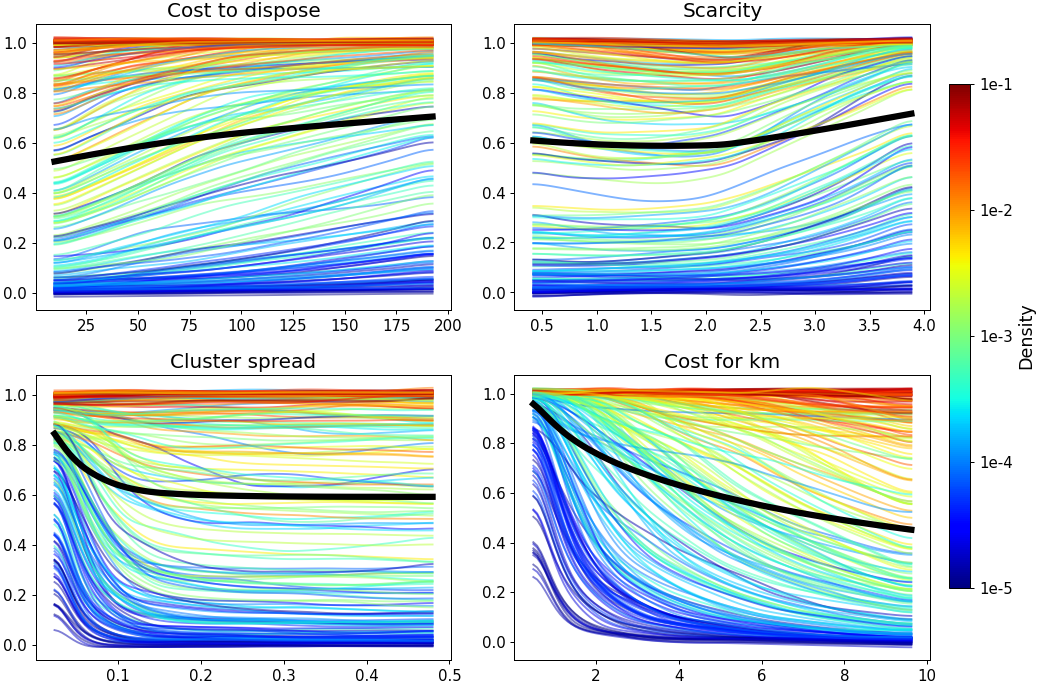}
        \caption{PDP and ICE illustrating the factors influencing the symbiosis index}
        \label{fig:pdp_symbiosis}
    \end{subfigure}
    \caption{PDP and ICE of $c_d,\ s, \ cd$ and $c_{t}$ colored by fixed densities for price and symbiosis.}
    \label{fig:pdpice}
    \vspace{-0.2cm} 
    \Description{Plots of the PDP and ICE of the disposal cost, the scarcity, \ cd$ and $c_{t}$ colored by fixed densities for price and symbiosis.}

\end{figure*}

Although firm density emerges as the most influential factor for both symbiosis and price, its overwhelming effect tends to dominate the variance and obscure the individual contributions of the other parameters. To better isolate and interpret the impact of the potentially actionable economic and spatial levers, we therefore fix density at different levels and focus the subsequent analysis on the remaining four input variables. 
Therefore, in the subsequent analysis, we focus on the 4 other input variables under different fixed density scenarios, since their interaction effects with density are all significant even at a third order. 
To better capture the non-linear effects, we replace the sparse PCE by a Multi-Layer Perceptron neural network surrogate model ($R^2$ of 97\% for the symbiosis and 95\% for the price), and use it to predict both the Partial Dependence Plots (PDP) and Individual Conditional Expectations (ICE)~\cite{robani2025smt} as illustrated in Figure~\ref{fig:pdpice}. In these plots, each colored line corresponds to an ICE, showing how the outcome of a single simulation responds when varying one parameter while holding others fixed; the black line is the PDP, representing the average effect across all simulations, and the color of each ICE line encodes the underlying firm density.

First, for the price in Figure~\ref{fig:pdp_price}, the PDP and ICE plots are consistent with the Sobol' analysis in Table~\ref{tab:sobol_symb_price_top}. In the blue low-density cases, the exchange price remains close to the market price and is only weakly influenced by other variables. As density increases, however, the price tends to decrease. The most influential variable is Scarcity, which exhibits a threshold effect: at low levels, scarcity has little impact, but between 1.5 and 2.5 it strongly drives the price upward. Beyond 2.5, the effect persists but grows at a slower rate. 
We also observe that increasing the disposal cost provides a strong and linear incentive for price reduction; the higher the density, the stronger the effect. Finally, cluster spread and kilometer cost have more limited influence, suggesting that they are less effective levers for intervention on the price outcome.

Second, for circularity in Figure~\ref{fig:pdp_symbiosis}, we again observe a pronounced density effect: higher density leads to greater symbiosis, largely independent of other variables.
At low density levels, cluster spread emerges as an important factor, but only when it takes on small values. As spread increases, its influence fades rapidly, suggesting that spatial compactness can partially compensate for low density but only within narrow limits.
Reducing travel costs appears more consistently actionable: at low density, for example, a per-kilometer cost below $2$ is required to reach any desirable symbiosis value. Under the same low density conditions, increasing disposal costs also provide a clear incentive to exchange byproducts, while at high density its effect is only marginal. Finally, scarcity has a notable effect on low-to-mid density values only above the critical threshold of 2.5.

In summary, firm density is a primary determinant of both symbiosis and price, though it is mostly exogenous and difficult to influence directly through policy. Among the controllable parameters, transport cost ($c_t$) and disposal penalty ($c_d$) emerge as effective levers. The sensitivity results indicate that reducing per-kilometer transport costs, through measures such as shared logistics or targeted subsidies, can substantially enhance exchange opportunities in low-density areas. Similarly, higher disposal penalties consistently incentivize firms to seek symbiotic exchanges instead of wasteful disposal.

\section{Discussion}
\label{sec:discussion}
Our results demonstrate that the proposed mechanism generates stable, economically grounded behaviors. Emergent pricing patterns reveal that sellers adapt efficiently to local conditions, while counterfactual regret analysis confirms convergence toward a near-Nash equilibrium rather than stochastic oscillation. Furthermore, global sensitivity analysis establishes clear causal links between model parameters and market outcomes. Significant high-order interactions suggest that circularity arises from the synergy of spatial configuration and economic incentives, highlighting the necessity for integrated, multivariate policy frameworks. Ultimately, these findings indicate that decentralized adaptive learning produces rational outcomes without centralized clearing, validating the mechanism's robustness.

Beyond methodology, the model offers a framework for evaluating policy instruments by mapping simulation parameters to regulatory levers. For instance, increasing disposal costs mimics landfill taxes, while the cluster spread parameter can be interpreted as a proxy for industrial zoning or policies that promote firm density.
This mapping enables the simulator to function as a virtual laboratory, allowing stakeholders to perform ``what-if'' analyzes and compare interventions before real-world deployment.

While these insights highlight the applicability of our model, we must recognize several limitations that temper the generalizability of the results.
We focused on a single byproduct to isolate the core mechanisms of decentralized exchange, abstracting away product heterogeneity and multi-market interdependence. 
Likewise, to ensure control and replicability, our experiments relied on stylized virtual geographies rather than empirically calibrated territories. 
We also abstract from temporal dynamics (production cycles, storage) and agent memory (trust, long-term contracts). 
Nonetheless, this work provides an indispensable first step toward systematically modeling spatially constrained markets for industrial symbiosis, upon which richer and more realistic frameworks can be built.
Future work will address these limitations by introducing multi-product environments, temporal dynamics, and agent memory to capture substitution effects and relational contracts. Grounding simulations in empirical case studies will further validate predictions and refine the policy decision tool. Finally, we also plan to include rigorous runtime benchmarking to formally showcase the simulator's scalability.

\section{Conclusion}
\label{sec:conclusion}

We introduced a decentralized multi-agent model for local byproduct exchanges, capturing how sellers adapt prices through RL, based on spatial and economic constraints. Simulations exhibit that adaptive decentralized strategies can produce stable circular outcomes.
Our results highlight two complementary contributions. 
First, from a \textit{policy design perspective}, the simulator is the initial step toward a decision support tool. By linking agent-level parameters to macro-level outcomes such as the symbiosis index, the model allows policymakers to identify which levers most strongly affect circularity. This provides a controlled environment where different regulatory scenarios can be tested before real-world implementation.
The emergent insights from our simulations help policymakers prioritize interventions that effectively promote circular practices across heterogeneous firms and contexts.
Second, from a \textit{MAS perspective}, the work provides a new methodological understanding of decentralized exchange mechanisms under spatial and economic constraints. We show how RL agents, interacting in a double auction market with heterogeneous resources and transport costs, can converge to stable trading patterns. 
Beyond the application domain, our contribution lies in revealing how market based coordination and learning dynamics shape emergent equilibrium.

\begin{acks}
This work was supported by the MUTTEC project (France 2030 – TIRIS, contract 23-AAP-TIRIS-01-062) and by the MIMICO project (ANR-24-CE23-0380).

\end{acks}

\bibliographystyle{ACM-Reference-Format} 
\bibliography{biblio}


\end{document}